\documentclass{ws-jai}
\usepackage[flushleft]{threeparttable}

\graphicspath{{./}{figures/}}

\usepackage{hyperref}
\usepackage{natbib}
\usepackage{float}
\usepackage{amsmath} 

\begin{document}

\catchline{}{}{}{}{} 

\markboth{Author's Name}{Paper Title} 

\title{Characterization and Quantum Efficiency Determination of Monocrystalline Silicon Solar Cells as Sensors for Precise Flux Calibration}

\author{
Sasha Brownsberger$^{1,2}$, 
Lige Zhang$^{2}$, 
David Andrade$^{3}$ and 
Christopher Stubbs$^{2,4}$}

\address{
$^{2}$Department of Physics, Cambridge, MA 02138, USA, sashabrownsberger@g.harvard.edu \\
$^{3}$School of Engineering and Applied Sciences Cambridge, MA 02138, USA\\
$^{4}$Department of Astronomy, Cambridge, MA 02138, USA, sashabrownsberger@g.harvard.edu 
}

\maketitle 

\corres{$^{1}$Corresponding Offer.}

\begin{history}
\received{(to be inserted by publisher)};
\revised{(to be inserted by publisher)};
\accepted{(to be inserted by publisher)};
\end{history}








\begin{abstract}

As the precision frontier of astrophysics advances towards the one millimagnitude level, flux calibration of photometric instrumentation remains an ongoing challenge.  
We present the results of a lab-bench assessment of the viability of monocrystalline silicon solar cells to serve as large-aperture (up to 125mm diameter), high-precision photodetectors.  
We measure the electrical properties, spatial response uniformity, quantum efficiency (QE), and frequency response of 3$^{rd}$ generation C60 solar cells, manufactured by Sunpower.   
Our new results, combined with our previous study of these cells' linearity, dark current, and noise characteristics, suggest that these devices hold considerable promise, with QE and linearity that rival those of traditional, small-aperture photodiodes. 
We argue that any photocalibration project that relies on precise knowledge of the intensity of a large-diameter optical beam should consider using solar cells as calibrating photodetectors.  
\end{abstract}



\section{Introduction} 
\label{sec:intro}

Projects at the frontier of precision astrophysics and cosmology, such as the Large Synoptic Survey of Space and Time (LSST) \cite{Ivezic2019}, require relative photometric calibration of 1\% to achieve their science goals.  When combined with precise distances to Galactic stars from the GAIA project \cite{GAIA2018}, it is also meaningful to strive for absolute calibration of flux at a comparable level. 
Reaching such ambitious calibration targets will be possible only if the imaging system's throughput, $T$, is known to better than 1\%.  
The LSST project's plan to measure $T$ is based on the methodology first laid out in \cite{Stubbs2006}. A monochromatic known dose of photons is injected into the telescope aperture and measured with both the telescope imaging system and with a precisely calibrated, reference photodetector.  

At present, the most promising arrangement to realize this \emph{in situ} calibration method uses a collimated beam projector (CBP), an instrument that emits a collimated monochromatic beam into the aperture of the telescope under study.  Successful CBP-style calibrations have already been performed by \citet{Coughlin2016, Coughlin2018}, and the conceptually similar StarDICE calibration project remains in active development \cite{Regnault2015}.  

The benefits of CBP-based calibration include:
\begin{itemize} 
\item direct control of spectral properties of the calibrating light, 
\item collimated CBP emission mimics the wavefront of collimated stellar emission, 
\item no need for atmospheric corrections, 
\item reduced susceptibility to systematic errors from stray light in the optical system. 
\end{itemize} 
However, the accuracy of CBP-based measurements of $T$ are no better than the accuracy to which the collimated beam is itself calibrated.  
Many CBP designs produce collimated beams with large cross sections (5cm in diameter or more), and calibrated such a large beam with traditional, small aperture photodiodes is difficult.  Optics that focus the collimated beam onto a photodiode themselves require calibration, and measuring only a section of the beam with a small photodiode would measure the true intensity of the full beam only if the beam were perfectly uniform.  

Motivated by this and other photo-calibration problems, we are engaged in an ongoing effort to test whether monocrystalline, back-contact silicon solar cells (SCs) could serve as large-aperture, high precision photodetectors.  The attractive attributes of these devices include: 
\begin{enumerate}
    \item large collecting area, spanning 125 $\times$ 125 mm, 
    \item large thickness, 165 $\mu$m, yielding good NIR quantum efficiency, 
    \item 100\% fill factor, with no electrodes on the light-facing surface, 
    \item high quantum (QE) efficiency, $>90\%$, across the spectral region of interest for instruments that use Si sensors, 
    \item affordability, 
    \item portability, allowing for relatively easy deployment in a variety of observatory settings. 
\end{enumerate}

We detailed our initial assessment of these devices, as well as the operating principle for back-contact monocrystalline SCs, in \citet{Paper1}. 
This paper describes progress in our ongoing effort to fully assess these devices' advantageous properties for precise flux characterization. 
Our work builds on previous high-fidelity characterizations of other SCs, tied to national-lab calibrated photodiodes \cite{Larason2008, Xu2012, Roller2019} .  
What sets our work (both \citet{Paper1} and this paper) apart is our specific and targeted focus on using characterized SCs to calibrate astronomical imaging systems.  We are principally focused on determining the SC properties that are necessary for sub-1\% determination of telescope throughput and designing SC containment systems that will allow the calibrated devices to be easily deployed on site.  

In Section \ref{sec:gens}, we discuss the differences between the previous generation and the new generation of studied SCs.  
In Section \ref{sec:RSchandSelect}, we provide measurements of the shunt resistances of 176 SCs, describe how we selected which devices warranted further characterization, and how we prepared selected devices for test. 
In Section \ref{sec:spatial}, we measure the small-scale spatial variations of the QE of the new SCs, noting that the effect is much reduced relative to the same behavior in the previous cell generation.
In Section \ref{sec:QE}, we show our 1\% measurement of the absolute QE of one mounted SC, and we describe the measurement setup that we used to arrive at our result.  
In Section \ref{sec:timeRes}, we determine that the frequency response of an SC is dominated by its intrinsic electrical characteristics (shunt resistance and shunt capacitance), and provide an example of measuring the RC time constant for an SC under test.  
We conclude in Section \ref{sec:conclusion} and discuss some expected short term applications of these calibrated SCs.

\section{C60 Solar Cell Generations} \label{sec:gens} 

Since we published our initial assessment of 2$^{nd}$ generation C60 SCs (Gen 2) in \citet{Paper1}, the Sunpower corporation released their improved 3$^{rd}$ generation C60 SCs (Gen 3).  
The Gen 3 devices differ from the Gen 2 devices in the following ways:  

\begin{itemize}
    \item More finely spaced backside electrode, now only 0.48 mm apart, compared to the 0.59 mm of Gen 2 SCs (see Figure \ref{fig:SCSpacing}).
    \item Improved AR coating, with less reflected blue light, giving the Gen 3 SCs good QE even into the ultraviolet (see Figure \ref{fig:QE}).   
    \item Different connection bond pads, with smaller footprints. 
    \item Systematically larger shunt capacitance and systematically smaller shunt resistance (see Figure \ref{fig:SCRShunt}), presumably due to tighter spacing of the alternating positive and negative electrodes. 
    \end{itemize}

We concluded that the new Gen 3 SCs are, on balance, better photometric calibrators than Gen 2 SCs.  
Where we show results of Gen 2 SCs, we explicitly say so.    

\section{The Solar Cell Shunt Resistances, Device Selection, and Device Preparation} \label{sec:RSchandSelect} 
\subsection{The Distribution of Solar Cell Shunt Resistances and Selection Criterion } \label{sec:Rsh} 
As we described in Section 3.2 of \citet{Paper1}, the ratio of an attached load's impedance, $R_L$, to the SC shunt resistance, $R_{sh}$, determines what fraction of SC photocurrent is discharged in the cell, going undetected by the load.  
Therefore, when assessing an individual SC's viability as a photometric calibrator, $R_{sh}$ is a vital figure of merit.  

\begin{figure}
   \centering
    \includegraphics[width=6in]{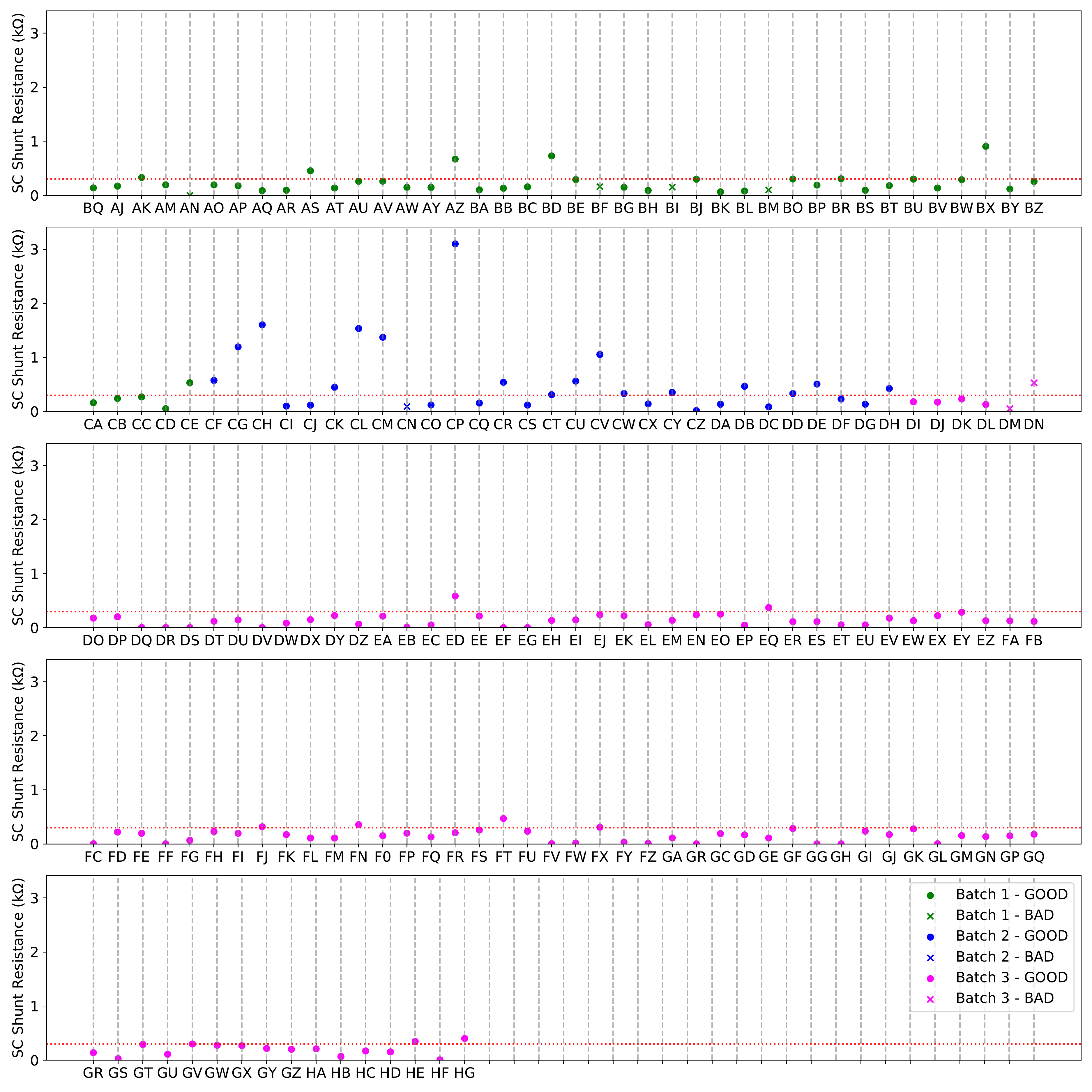}
    \caption{The distribution of shunt resistances, $R_{sh}$, of Gen 2 (batch 1 and 2) and Gen 3 (batch 3) C60 solar cells (SCs), 176 in total.  Note that the third generation SCs (magenta) have systematically smaller shunt resistances.  The horizontal line at $R_{sh} = 300 \Omega$ defined our cut off for cell selection.  Those SCs marked `BAD' were damaged either before or during characterization.}
    \label{fig:SCRShunt}
 \end{figure}

To better understand the variance of this important detector property in mass-produced batches of C60 SCs, we measured $R_{sh}$ of 176 SCs from both Gen 2 and Gen 3. 
We performed this measurement using the basic method first implemented by \citet{Chen1984}, by covering the front and back of the SC under test with black foil and applying an ohmmeter to its external wire bond mount points.  We show each of the 176 measured $R_{sh}$ values in Figure \ref{fig:SCRShunt}.  Cell IDs were assigned in order in which $R_{sh}$ was measured. 

We found that the Gen 3 SCs had systematically lower $R_{sh}$ than the Gen 2 SCs, presumably because the electrodes are more tightly packed, facilitating electron leakage between electrodes of the sort described by \citet{Banerjee1986}.  
In all SC batches that we studied, most devices had unacceptably small $R_{sh}$ values.  Setting our selection criterion at $R_{sh} > 300 \Omega$, 30\% of Gen 2 cells and 8\% of Gen 3 cells were acceptable. 
We selected only SCs that met this selection criterion for further study.  The results in this paper are mostly based on measurements of Gen 3 solar cell ED.

\subsection{Device Preparation} \label{sec:prep} 

\begin{figure}[htp]
    \centering
    \includegraphics[width=5in]{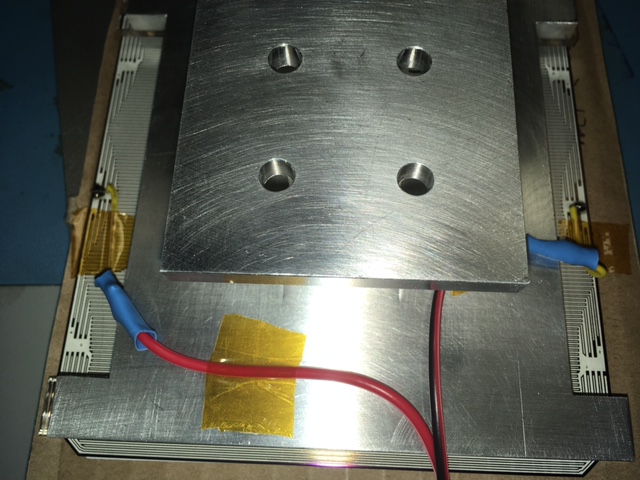}
    \caption{Electrical connection and packaging of a Gen 3 C60 solar cell. A piece of solid wire is temporarily held in place by Kapton tape, and is soldered to a larger gauge BNC pigtail.  The solar cell, facing down in this image, is glued to an insulating plexiglass spacer that is in turn glued to the aluminum frame shown. This frame is then mounted inside the detector housing shown in Figure \ref{fig:Gen3Packaged}. The temporary Kapton tape is removed before operation.}
    \label{fig:Gen3Electrical}
\end{figure}

 After selecting SCs with acceptably large shunt resistances (see Section \ref{sec:Rsh}), we experimented with various soldering methods and electrically conductive adhesives to connect wire leads to the electrodes of an SC.  We settled on a method of soldering a solid piece of 24-gauge wire to the central solar cell electrode, and then connecting a pigtail BNC connector to a ground-isolated bulkhead BNC feed-through connector.  This electrical connection is shown in Figure \ref{fig:Gen3Electrical}.  This method produces more stable electric bonds than adhesive-based connection methods. 
 
\begin{figure}[htp]
    \centering
    \includegraphics[width=5in]{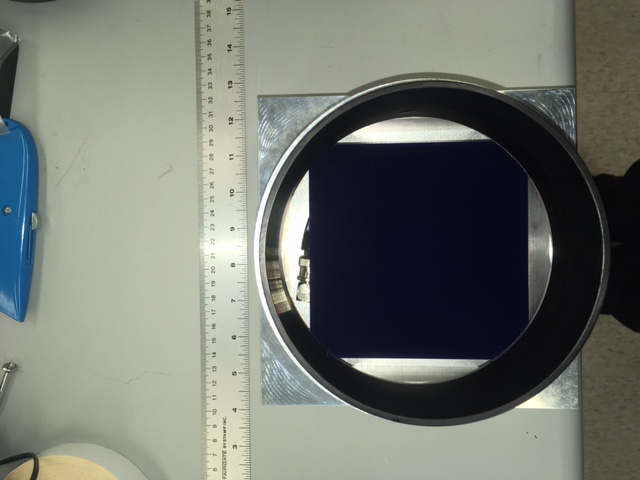}
    \caption{Image of packaged solar cell. An upstream aperture limits illumination to the central area of the solar cell. }
    \label{fig:Gen3Packaged}
\end{figure}

 We designed specialized housing to hold a single SC under study.  Characteristics of these housings include: 
 \begin{itemize} 
 \item a robust, metal casing to protect the delicate SC, 
 \item a removable front pane to make inserting the SC easy,
 \item a metal snout extending 5 cm in front of the cell to block background light contamination, 
 \item internal 1/4"-20 threads to mount the SC casing to optical stands, 
 \item a mounted BNC connector so that the SC wire leads can be connected to external electronics.
 \end{itemize} 
To mount an SC in this housing, the device was first glued to a piece of electrically insulating plexiglass, which was then glued to the interior of the aluminum SC casing.  We show an image of the fully packaged cell in Figure \ref{fig:Gen3Packaged}. 

\section{Spatial Variations in Solar Cell Quantum Efficiency}  \label{sec:spatial} 
\begin{figure}
    \centering
    \includegraphics[width=6in]{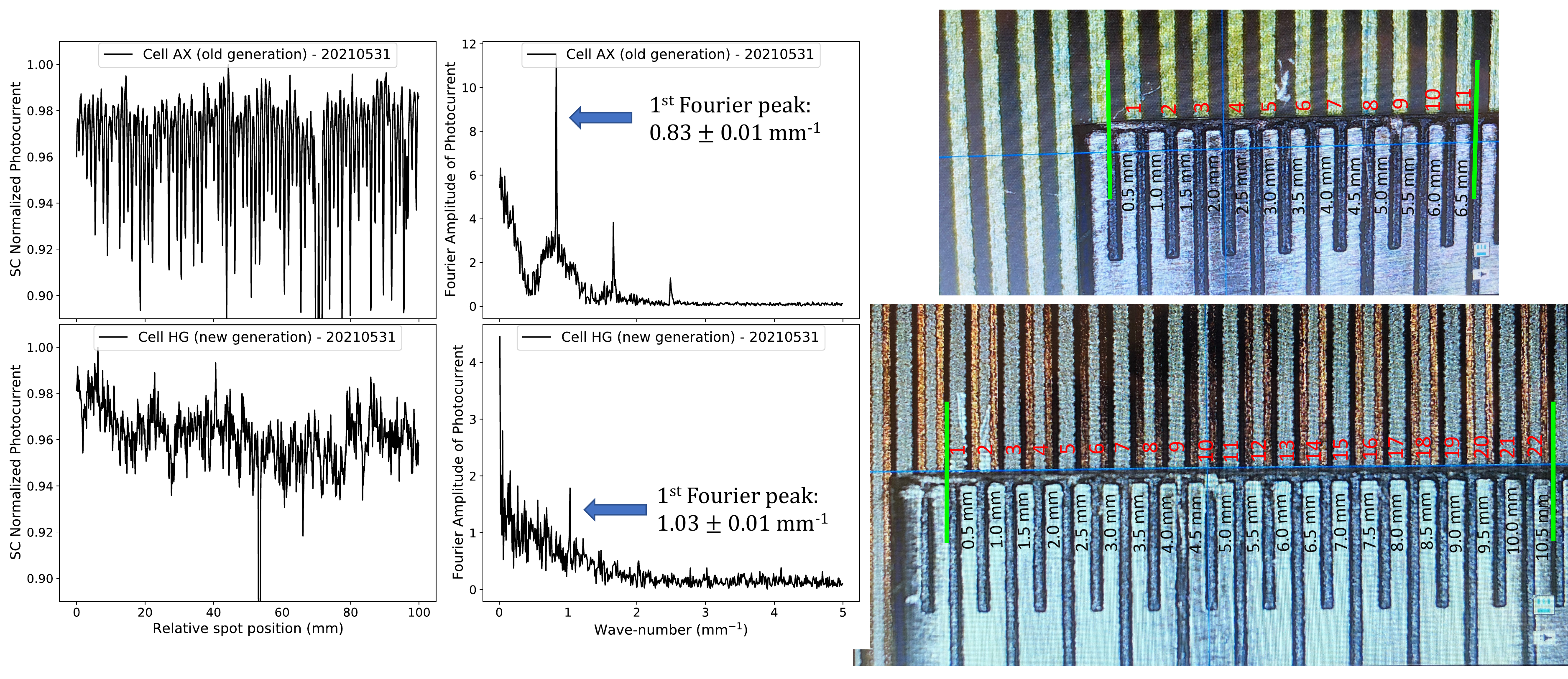}
    \caption{ The normalized photocurrent as we move a 0.1mm spot of light across the front of a solar cell orthogonal to the backside electrodes (left column), the Fourier amplitude of that normalized photocurrent (middle column), and a magnified image of the back-side spatial structure with a ruler included for scale (right column) of a Gen 2 (top row) and Gen 3 (bottom row) SunPower C60 solar cell.  The locations of the first Fourier peaks of the Gen 2 and Gen 3 SCs are respectively at wavenumbers of $0.83 \pm 0.01$mm$^{-1}$ and $1.03\pm 0.01$mm$^{-1}$.  
    These measurements are consistent with the measured spacing of the backside electrodes: in the Gen 2 and Gen 3 cells, we respectively measure electrode spacings of $0.58 \pm 0.01 $mm and $0.48\pm 0.01$ mm, corresponding to same-voltage electrode spacings of $1.18 \pm 0.01$mm and $0.96 \pm 0.01$mm.  As we described in \citet{Paper1}, the alternating voltages of the back contact electrodes produce a corrugated electric field, resulting in charge traps that reduce the cells' effective QE.  This charge trapping effect is a factor of five less pronounced in the Gen 3 SCs, likely due to their tighter electrode spacing and a reduction in the relative size of differently charged electrodes.}
    \label{fig:SCSpacing}
 \end{figure}

In \citet{Paper1}, we noticed a 10\% variation in the QE of a Gen 2 SC as we moved a small spot of light across the cell's front face, and we noted that such a substantial change in SC QE would introduce significant systematic error to any calibration effort involving small diameter light beams.  
These QE variations are due to the charge traps produced by the corrugated electric fields of the alternating positive and negative SC electrodes.  
Because the electrodes are more tightly spaced in Gen 3 SCs, we repeat the QE response measurements, scanning a 100$\mu$m spot across the surface of Gen 3 cell ED.  We show the results of both the old Gen 2 scan and this new Gen 3 scan in Figure \ref{fig:SCSpacing}. 

The Gen 2 and Gen 3 same-polarity electrodes are respectively spaced $1.18\pm 0.01$ mm and $0.96\pm 0.01$mm apart.  Consistent with these spacings, we find that the largest Fourier peaks of the spatial scans are located at wavenumbers of $0.83 \pm 0.01$mm$^{-1}$ and $1.03\pm0.01$mm$^{-1}$.  The size of the Fourier peak in the Gen 3 cell is only 20\% the height of the Fourier peak in the Gen 2. cell, relative to the continuum.  
Owing to the tighter spacing of their conducting electrodes, the QE of the Gen 3 cells varies by only 2\% on small spatial scales.  Solar cell flux measurement errors due to spatial QE variations will therefore be smaller in Gen 3 cells.   

\section{Absolute Quantum Efficiency Determination} \label{sec:QE} 
\subsection{Method and setup} \label{sec:QEMethod} 

\begin{figure}[htp]
    \centering
    \includegraphics[width=6in]{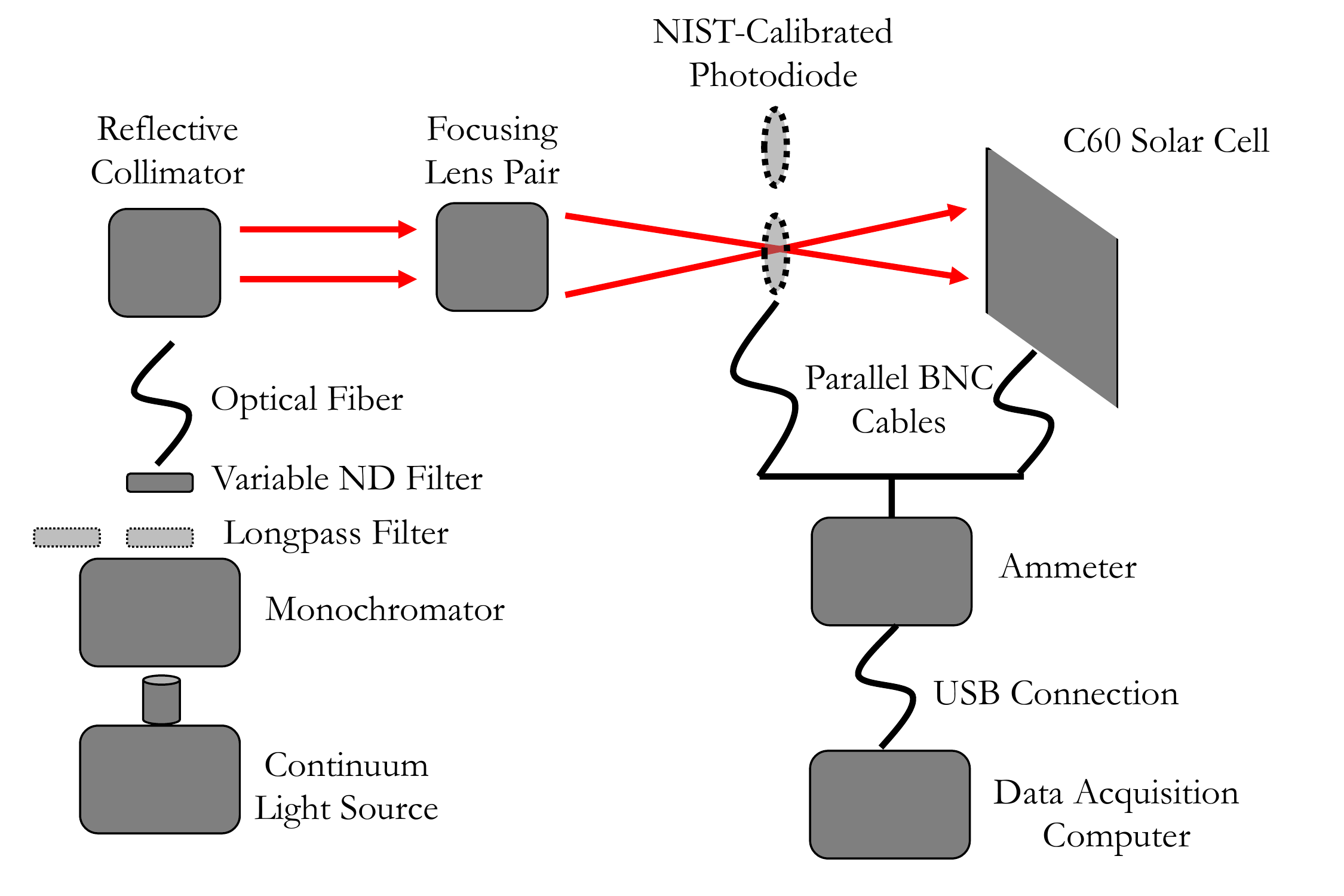}
    \caption{The experimental setup used to tie the quantum efficiency (QE) of a solar cell (SC) under test to that of a NIST-calibrated photodiode (PD).  The monochromator, with an optical bandwidth of 5nm, illuminates a variable ND filter and then the tip of an optical fiber.  The optical fiber feeds a reflective collimator, the collimated output of which is focused by a pair of lenses.  
    We place the PD on a mechanical flipper, allowing us to remotely move it into and out of the waste of the focused beam.  
    The SC rests at the end of this optical chain, with $\simeq$50\% of its photoconducting surface illuminated by the monochromatic light when the PD is flipped out of the beam path. 
    Both SC and PD are wired in parallel, their generated photocurrents aggregated and fed to a single high-fidelity current meter.  
    A Windows laptop controls the moving components of the measurement setup and stores the acquired data.  
        We mount the variable ND filter to a digitally controlled translation stage, allowing us to adjust the photodiode photocurrent to a target value of 0.9$\mu$A over most of the wavelength range (see the upper left panel of Figure \ref{fig:QE}).  
        For wavelengths above 650nm, we insert a 590nm longpass filter in the beam path to block second order light contamination.   Not shown are baffles and stray light controls that are essential to arrive at the desired accuracy. }
    \label{fig:QElayout}
\end{figure}

We designed an experimental setup to measure the absolute quantum efficiency (QE) of a single C60 SC to better than 1\% accuracy by tying the measured SC photocurrent to that of a NIST-calibrated \cite{Larason2008} photodiode (PD).  
Figure \ref{fig:QElayout} shows our experimental setup.  
The PD is placed at the waist of a diverging beam on a mechanical flipper, and the SC is placed further back in the beam path.  We position both the PD and SC so that the beam underfills their photosensitive areas.  

To avoid systematic errors associated with the readout signal chain, we wire both detectors in parallel and feed their aggregate signals to the same series ammeter and parallel voltmeter. 
An order-blocking filter eliminates second order light contamination when operating at wavelengths longer than 650 nm. The variable ND filter is used to adjust the photocurrent in the calibrated photodiode to a target of 0.9$\mu$A when possible.  

Table \ref{tab:sometab} lists the details and part numbers of the devices used.

\begin{table}
  \centering
  \begin{tabular}{|c|c|c|}
    \hline
Part & Manufacturer & Product Number \\
\hline
Continuum light source & Thorlabs & SLS302 \\ 
Monochromator & Newport & Corner Stone 260   \\ 
Longpass filter & Edmund Optics & 46 063 OG590  \\ 
Variable ND Filter &  Edmund Optics & 41960 \\ 
Optical Fiber & Thorlabs & M93L02 \\ 
Reflective Collimator & Thorlabs & RC08SMA-F01 \\
Focusing lenses (2) & Thorlabs  & LA1050-A-ML \& LA1417-A-ML \\ 
Photodiode & Hamamatsu, NIST calibrated & S2281 \\ 
Solar cell & SunPower & C60 \\ 
Cell enclosure & Custom  & N/A \\ 
Ammeter & Keysight & B2987A \\ 
\hline
  \end{tabular}
  \caption{Table 1 - Components of the setup used to measure the QE of a C60 solar cell using a NIST calibrated Hamamatsu photodiode shown in Figure \ref{fig:QElayout}. } \label{tab:sometab}
\end{table}

\subsection{Results} \label{sec:QERes}
To measure the QE of the solar cell, we repeatedly scan the monochromator from 350-1050nm, in 1nm steps.  At each nm, we measure the photocurrent with the photodiode in the beam path with the monochromator shutter open and closed, then we flipped the photodiode out of the beam path and measure the photocurrent in the SC with the shutter open and closed.  Each photocurrent measurement consists of 200 individual current readings, at a sampling frequency of 60 Hz.   
To measure the photocurrent in the SC and the PD, we discard the first 50 components of each sample chain to avoid shutter and flipper artifacts and then measure the median of the remaining samples.  
The median dark current (shutter closed) value is subtracted from the corresponding shutter open median value.   
Running this measurement continuously for approximately three days, we densely and repeatedly scan the SC QE curve.   

\begin{figure}[htp]
    \centering
    \includegraphics[width=6in]{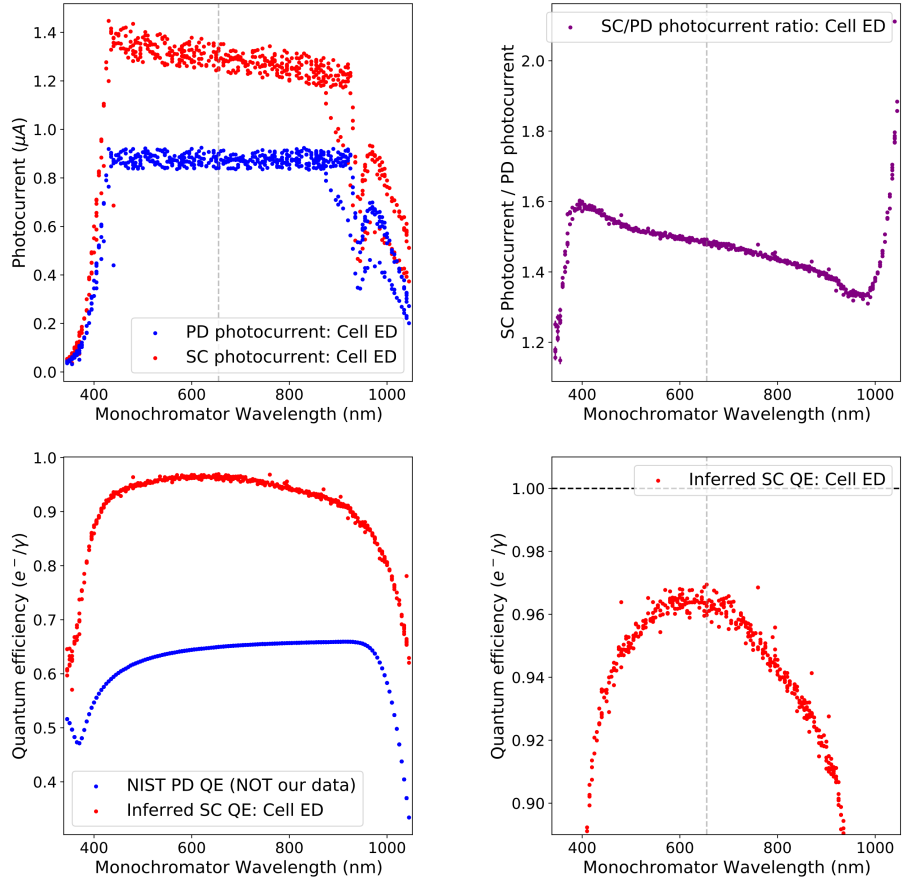}
    \caption{The data sets from which we measure the quantum efficiency (QE) of one Gen 3 C60 solar cell (SC) ED.  Starting with the photocurrents in the NIST-calibrated photodiode (PD) and SC ED (upper left), we take the ratio of the photocurrents (upper right) and divide that ratio by the reported QE of the PD (lower left plot, blue curve) to acquire our measured QE of cell ED (lower left plot, red curve).  We show the high-QE section of the cell ED's measured QE curve in the lower right panel.  
    The SC has excellent QE, peaking at 96\% and staying above 90\% between about 400nm and 900nm.  
    The scatter in the measured points was due principally to instability in the SC dark current and small drifts in the continuum source intensity.  Even with these systematic uncertainties, our measured SC QE is stable to better than 1\%. }
    \label{fig:QE}
\end{figure}

We show the results of this long scan in Figure \ref{fig:QE}. 
The C60 SC that we study (cell ED) has high QE over the spectral range of primary interest - greater than $90\%$ between 450nm and 950nm, and above $70\%$ between 400 and 1000nm.  Relative to the NIST-calibrated PD, the SC's QE is particularly high between 375 and 400nm, where the effect of the improved AR coating is made especially clear. 
Our results are consistent between scans, showing no drift.  
We report the absolute QE of a C60 solar cell to better than $1\%$.

\section{Bandwidth and Frequency Response} \label{sec:timeRes} 

\begin{figure}[htp]
    \centering
    \includegraphics[width=6in]{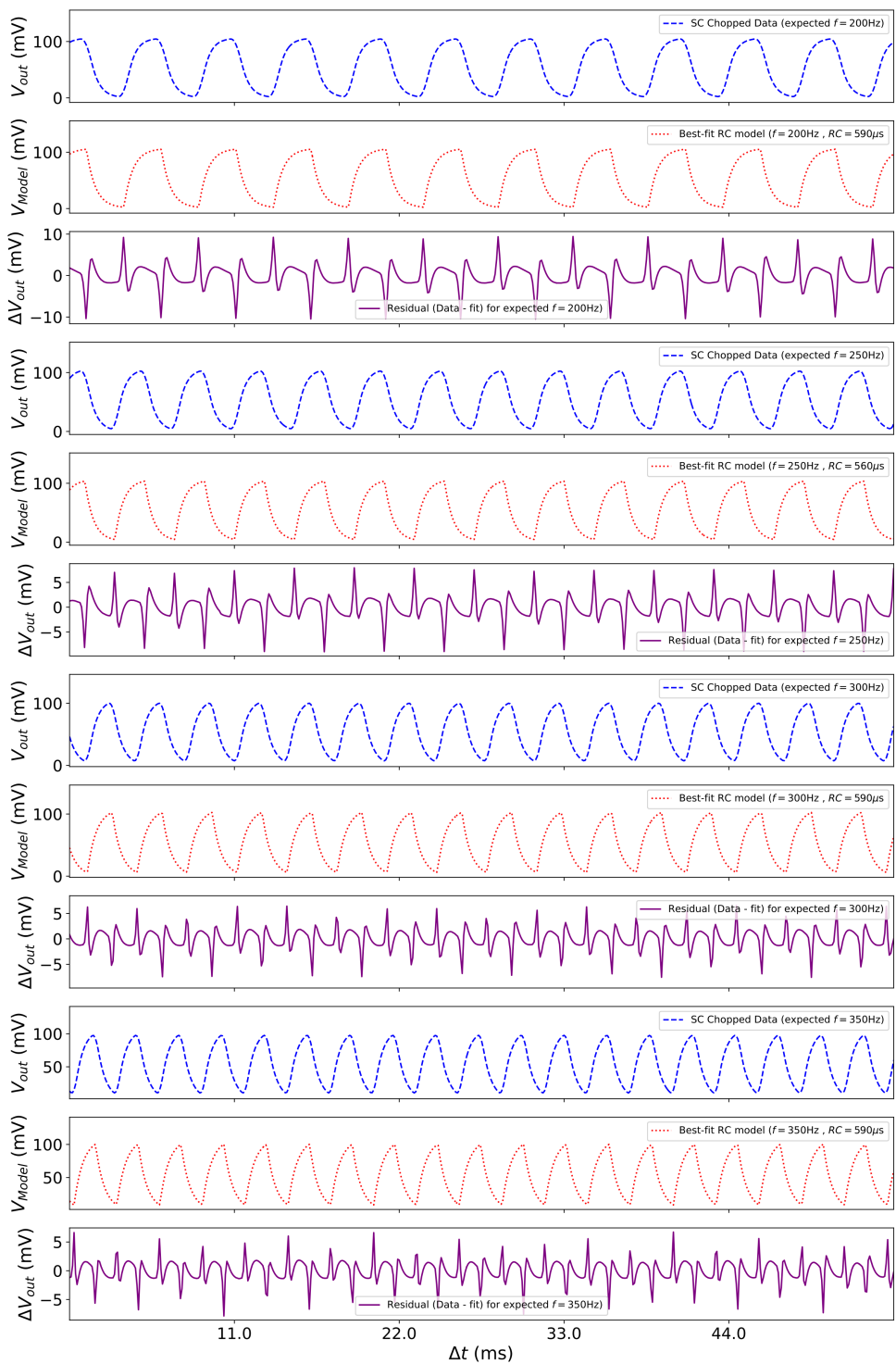}
    \caption{The time response of the C60 solar cell ED to chopped illumination (blue, dashed), the best-fit RC circuit response (red, dotted), and the difference between the data and model (solid, purple) at increasing frequencies.     
    An RC model with a time constant of 550-600$\mu$s characterized the SC time response curve well, though not perfectly.  To first order, the time response of a C60 solar cell is dominated by the internal shunt capacitance and resistance of the cell.  }
    \label{fig:RCResponse}
\end{figure}

To understand the effect that a possibly slow photoelectric response and a relatively large internal shunt capacitance might have on our ability to use chopped illumination with SC-based photocalibration, we re-purpose the QE setup described in Section \ref{sec:QEMethod} to measure the time-response of solar cell ED.   We change the setup in three ways: 
\begin{itemize} 
\item switch from the monochromator to a brighter, broad-band light source, 
\item place an optical chopper in the path of the converging beam, before the calibrated photodiode,  
\item exchange the ammeter with a voltmeter, inferring photocurrent response by measuring the voltage as the photocurrent was discharged over the SC internal shunt resistance, $R_{sh}$. 
\end{itemize} 
Running the chopper wheel at various frequencies, $f$, we measure the time-varying voltage induced across cell ED, in photovoltaic mode.  Under the assumption that the RC shunt circuit of the C60 solar cell dominates the SC time response, we fit the measured output voltage with the standard RC `sharkfin' waveform.  The results are shown in Figure \ref{fig:RCResponse}.  

We find that, for chopper frequencies above 200Hz, an RC waveform with a time constant of 580$\mu$s effectively characterizes the photocurrent-induced voltage.  This suggests that the time response of a C60 solar cell is dominated by the internal electronic properties of the cell itself.  
     For photocalibration purposes, the strobe period of a calibration light source should be at least five times longer than the RC time constant.  
     For cell ED with an RC value of 580$\mu$s, this implies a strobe frequency of $<1/ (5 \times 580 \mu \textrm{s}) \simeq  340$Hz. This is comfortably above both power line harmonics and 1/f noise corners for well-chosen electronics. 

\section{Conclusion and a Possible Photocalibration Application } \label{sec:conclusion} 
We have continued our assessment of monocrystalline back-contact silicon solar cells (SCs) as possible high-precision photometric calibrators, begun in \citet{Paper1}.  

In this new work, we discussed the differences between 2$^{nd}$ generation (Gen 2) and 3$^{rd}$ generation (Gen 3) SCs (Section \ref{sec:gens}), and argued that Gen 3 cells should be used in calibration efforts.  
Following our conclusion in \citet{Paper1} that the internal SC shunt resistance, $R_{sh}$, is a vital figure of merit when assessing these devices' viability as photocalibrators, we measured $R_{sh}$ for 176 SCs of both generations.  
Setting our minimum acceptable $R_{sh}$ vale at 300$\Omega$, we found that about 30\% of Gen 2 cells and 8\% of Gen 3. cells could be used in further calibration efforts.  
We detailed our improved method for mounting these devices in Section \ref{sec:prep}.  

Because Gen 3 cells have more tightly spaced back-contact electrodes than Gen 2 cells, we repeated in Section \ref{sec:spatial} the small-scale QE variation measurement that we developed in \cite{Paper1}.  
Consistent with our expectations, we found that the Gen 3 cells had more rapidly varying and smaller amplitude QE variations than the Gen 2 cells.  For photo-calibration purposes, this smaller spatial variation will introduce less systematic error into flux measurements.     

For the first time in our analysis of these SCs, we devised and implemented a measurement to tie the QE of a SC under study to the QE of a NIST-calibrated photodiode (Section \ref{sec:QEMethod}). 
Using this setup for several uninterrupted days, we determined the absolute QE of one SC to better than 1\% precision.  
We found that these SCs exhibit excellent QE, peaking around 96\% at 600nm and remaining above 90\%  between 400 and 900nm.  We also noted that the Gen 3 cells have particularly good QE in the blue, relative to the reference PD, presumably owing to their improved AR coating.  

Considering our intention to use SCs as photometric calibrators with chopped light sources, we measured in Section \ref{sec:timeRes} the frequency-response of a Gen 3 SC.  We determined that a SC's time response is dominated by its internal electronic properties (shunt resistance and shunt capacitance).   In particular, we found that the time-response of cell ED to a chopped beam was well described by a standard RC waveform, with a time constant of 580$\mu$s.  As with the shunt resistances of the individual SCs (Figure \ref{fig:SCRShunt}), we fully expect that the distribution of RC time constants will vary across cells.  Each cell that is slated for use in photocalibration should similarly have its RC time constant characterized. 

  Considering the combined results of our initial analysis in \citet{Paper1} and those we present here, we are confident that solar cells are one of the best, and perhaps the best, tools available for achieving 1\% precision measurements of large-diameter optical beams.  
  We are actively using and intend to use the SCs that we have characterized in various high-precision calibration projects, including the full characterization of the StarDICE determination of telescope throughput \cite{Regnault2015} already underway. 
  
   The next generation of advanced telescopes demands exquisite calibration to realize their ambitious science goals.  
   The use of terrestrial, large-diameter light sources, built to mimic collimated starlight, will play a crucial in these calibration efforts.  
   We expect that large-area, high-fidelity photodetectors, such as the C60 SCs that we have studied, will play an important role in accurately characterizing such calibration beams.  

\section*{Acknowledgments}
We are grateful to the US Department of Energy for their support of our optimization and precision calibration efforts for the Rubin Observatory, under DOE grant DE-SC0007881, and to Harvard University for its support of our program.

\bibliography{SolarCells.bib}



\end{document}